# Unified Modelling of Infrastructure Asset Performance Deterioration – a bounded gamma process approach

Wang Chen and Arnold X.-X. Yuan*

*Department of Civil Engineering, Toronto Metropolitan University, 350 Victoria Street, Toronto, Ontario, Canada M5B 2K3.*

Corresponding author: Arnold X.-X. Yuan, E-mail address: arnold.yuan@torontomu.ca

# Unified Modelling of Infrastructure Asset Performance Deterioration –

# a bounded gamma process approach

**Abstract**

Infrastructure asset management systems require a flexible deterioration model that can handle various degradation patterns in a unified way. Owing to its appealing monotonic sample paths, independent increments and mathematical tractability, gamma process has been widely employed as an infrastructure performance deterioration model. This model was recently enhanced by introducing an upper bound to satisfy a practical modelling need that many infrastructure performance deterioration processes are constrained by physical or managerial limits. Several bounded transformed gamma process (BTGP) alternatives had been proposed; however, they lacked due flexibility to characterize different deterioration patterns. This paper proposed a new BTGP model that is deeply grounded upon the traditional regression modelling tradition in infrastructure asset management systems. Qualitative and quantitative comparisons were carried out between the proposed BTGP and a bounded nonstationary gamma process (BNGP) model from both deterioration modelling and asset management decision-making perspectives. An empirical study using the real-world historical bridge condition data was performed to examine the flexibility of the BTGP against the BNGP and six other BTGP alternatives. The results confirmed the flexibility and significance of the proposed BTGP model for infrastructure systems.

# 1 Introduction

Physical infrastructure systems (e.g., bridges, buildings, roads, water mains, sewer pipes, nuclear piping) and equipment all degrade with time and under extreme events. As their service age accumulates, the system performance deteriorates from an intact state until failure, when rehabilitation or reconstruction has to be performed. For greater efficiency, maintenance and repair activities may be implemented preventively before the system reaches a very poor or failed state. Therefore, characterizing the trend of performance deterioration, or deterioration modelling, becomes a critical analytical component for infrastructure asset management planning (Alaswad & Xiang, 2017; Prakash et al., 2020; Wu et al., 2021). Once the deterioration model is characterized and estimated, the future performance and remaining life of the infrastructure system can be predicted, and a proper preventive condition-based maintenance plan can be developed and implemented.

Deterioration modelling requires not only the quantification of the mean curve along time, but also the characterization and quantification of various uncertainties, among which temporal uncertainty – variability of the deterioration rate with time – is a critical consideration (Yuan, 2017). For this purpose, many stochastic process models have been developed. The discrete-state Markov chain is one of the most popular models (Calvert et al., 2020; Daneshkhah et al., 2017; Lin et al., 2019; Mizutani & Yuan, 2023). For continuous-state deterioration phenomena, the gamma process model remains the most popular (Bismut et al., 2022; Bismut & Straub, 2021; Liu et al., 2021; Pandey et al., 2009; van Noortwijk, 2009; van Noortwijk & Frangopol, 2004) among others including the inverse Gaussian process (Zhang & Zhou, 2014; Zhu & Hao, 2021) and geometric Brownian motion (Mizutani et al., 2024; Zhang & Zhou, 2015).

Due to its monotonicity and mathematical tractability, the gamma process model has been widely applied in the lifecycle decision support of many systems with complex inspection and maintenance needs across various engineering domains, including nuclear systems (Yuan, 2007), buildings (Edirisinghe et al., 2013), and pipes (Mahmoodian & Alani, 2014). It has been applied across various settings, including systems with dependent degradation processes (Bautista et al., 2022; Zheng et al., 2024), with random effects (Esposito et al., 2023), and with mixed stochastic degradation process (Dai et al., 2022; Li et al., 2023), and inspection planning of infrastructure system considering temporal and parameter uncertainties (Yuan et al., 2021). Further methodological developments have also emerged to accelerate degradation analysis to consider time-varying stress or usage conditions (Ling et al., 2015).

A major drawback of the conventional gamma process is that the sample paths as well as the mean and variance functions are unbounded (Guida et al., 2012). This does not match most of the degradation phenomena in reality. For example, flow-induced wall thinning of nuclear piping in structural integrity management always has a dimensional limit (Deng & Pandey, 2017; Yuan et al., 2008). In infrastructure asset management, infrastructure performance measures such as bridge condition index, pavement condition index, and facility condition index often range from 0 to 10 or 100. Without interventions such as maintenance, repair and rehabilitation, the performance will surely drop to nearly zero after a sufficiently long period. In addition, researchers have studied several empirical datasets and found an interesting phenomenon accompanied with the bounded degradation processes; that is, the variance of degradation first increases and then decreases over time (Giorgio et al., 2011; Giorgio & Pulcini, 2024; Guida & Pulcini, 2011).

To characterize the bounded degradation paths as well as the non-monotonic variance pattern, several bounded gamma process models were proposed. For example, Pulcini (2001) proposed a nonhomogeneous Poisson process with bounded failure intensity to fit the failure data of degrading systems exhibiting a bounded intensity function. Giorgio et al. (2015) proposed a new transformed gamma process with a mathematically tractable formulation, which accommodates age- and state-dependent degradation models. Deng and Pandey (2017) followed this idea and examined its application to model a corrosion-induced wall thinning process. After that, Giorgio and Pulcini (2018) further proposed another new transformed process model with a gamma-distributed scale parameter for the underlying gamma process and studied the consequences of model misspecification through a maintenance decision case of laser equipment. Most recently, Fouladirad et al. (2023) and Giorgio and Pulcini (2024) provided systematic comparisons of various bounded transform schemes and studied "model misspecification" issues from both a performance and lifetime prediction perspective and a maintenance decision-making perspective. All the transformed process models follow the same structure that involves two components: an underlying gamma process and a transform function.

While the bounded transformed gamma processes (BTGP) seem to be natural candidates for modelling various deterioration phenomena facing in infrastructure asset management, a problem of the existing BTGP models arises; that is the lack of flexibility. These models all have an underling nonstationary gamma process incorporated with a fairly rigid transform function. In a most recent work, Giorgio and Pulcini (2024) included an additional parameter in the transform function, which improves the maximized likelihood; however, the Akaike Information Criterion (AIC) value does not necessarily decrease for some cases. This may limit the modelling flexibility in fitting real-life degradation patterns.

Asset performance deterioration models in most current infrastructure asset management systems can be divided into two types. They are either a traditional regression model or a Markov chain model. A regression model fits a continuous-state deterioration phenomenon very well. With a nonlinear mean function of time, they are so flexible enough to fit different deterioration patterns (exponential decay, S-shape, etc). However, it is disadvantageous for its improper handling of temporal uncertainties (Yuan, 2017), even when a mixed-effects approach is taken (Yuan & Pandey, 2009). On contrast, Markov chain model can handle temporal variability very well; however, it works only for discrete-state deterioration processes.

Therefore, this paper aims to develop a flexible continuous-state deterioration model that is not only stochastic, mathematical tractable and bounded, but also flexible enough to adapt to various degradation trends. Model flexibility is an important criterion for stochastic deterioration model selection, because a good infrastructure asset management system is expected to handle various asset classes using a unified modelling approach. In asset management system design, it is not a good idea to offer a large number of options and leave it to the end users to choose the best one. This is not only because it would add a large computational burden, but also because end users often do not have the due technical capacity to choose the best model.

We took a straightforward model construction approach. Using a stochastic inner clock concept, we were able to integrate the traditional regression model into the modern stochastic process modelling framework. An intuitive interpretation was offered for the new BTGP model. We used both simulated and real-life data to test the model flexibility. To establish the necessity of such a bounded process model, we also designed a bounded non-stationary gamma process (BNGP) with a bounded shape function so that both the mean and variance functions have an upper limit, making the comparison between the conventional gamma process and the proposed BTGP fairer than the comparisons reported in literature.

In the following, we introduce the two competing models and their major properties. Their probabilistic characteristics are also summarized and compared, with particular attention given to a critical parameter – the mean value at maximum variance. Then, the flexibility of the proposed model against the BNGP and other BTGP alternatives are examined using real-life historical bridge condition data. After that, we study the impacts of the two competing models on age- and condition-based replacement policies. The last section concludes the paper with major findings.

## 2 Model Description and Characterizations

### 2.1 Conventional gamma process

This subsection reviews the very basics of gamma process and introduces the bounded nonstationary gamma process (BNGP). It is included for the ease of reference. The word 'conventional' is used here to distinguish the transformed gamma process to be introduced in the next subsection. A conventional gamma process, denoted as $X(t)$, is a stochastic process with continuous and monotonically increasing sample paths, always starting with zero at time 0. More importantly, the gamma process has independent increments; that is, the two increments of the process over any two non-overlapping time intervals are stochastically independent of each other. This suggests that the gamma process is Markovian. Actually, the independent increments property is stronger than a Markovian property. Of course, the defining feature of the gamma process is its gamma distribution property: At any given time $t$, the process state $X(t)$ is a gamma distributed random variable. More precisely,

$$f_{X(t)}(x) = \frac{x^{\alpha(t)-1}}{\beta^{\alpha(t)}\Gamma[\alpha(t)]} e^{-x/\beta}, \tag{1}$$

in which $\alpha(t)$ is referred to as the shape parameter, and $\beta$ as the scale parameter. Since the shape parameter is a function of time $t$, it is also called a shape function. The shape function must be a nonnegative, monotonically increasing function, and the scale parameter must be greater than 0. Note that gamma distribution in general is infinitely divisible. As a result, any increment of the gamma process between $t$ and $t + \Delta t$ is gamma distributed with shape $\Delta \alpha_t = \alpha(t + \Delta t) - \alpha(t)$ and the same scale $\beta$.

When the shape function is a linear function of time, the gamma process is called a stationary gamma process; otherwise, nonstationary. The stationary process has another feature; that is, increments over any non-overlapping time intervals of an equal duration are independent and identically distributed (i.i.d.) random variables. This feature is particularly appealing for deterioration modelling when the inspection intervals are fixed, as this turns a stochastic process modelling problem into one of random variable modelling. When the scale parameter $\beta = 1$, the stationary gamma process is also referred to as the standard gamma process.

The mean and variance of a conventional gamma process are expressed as:

$$\text{E}[X(t)] = \beta \alpha(t), \qquad \text{Var}[X(t)] = \beta^2 \alpha(t). \tag{2}$$

Therefore, the coefficient of variation of the gamma process is independent of the scale parameter. For a stationary gamma process, both the mean and variance are a linear function of time.

To account for the boundedness of degradation process in reality, one approach is to introduce a bounded shape function, which can at least characterize the feature in the mean sense, as Equation (2) suggests. A flexible option is the following:

$$\alpha(t) = \frac{x_{lim}}{\beta}\left(1 - \exp\left[-\left(\frac{t}{\theta_3}\right)^{\theta_2}\right]\right), \tag{3}$$

in which $x_{lim}$ represents the upper limit of the gamma process, $\theta_2$ and $\theta_3$ are two parameters of the shape function. For the ease of presentation later, we reserve $\theta_1$ for the scale parameter $\beta$ of the nonstationary gamma process. The expression inside the bracket is the same as the cumulative distribution function of a Weibull distribution. Different combinations of $\theta_2$ and $\theta_3$ offer different growth patterns of the non-stationary gamma process. Substituting Equation (3) into (2), one can readily show that as $t$ goes to infinity, the mean value and standard deviation of $X(t)$ converges to $x_{lim}$ and $\sqrt{\beta x_{lim}}$, respectively. This process with bounded mean path is hereafter referred to as the bounded nonstationary gamma process (BNGP), or simply the BN model.

It must be noted that although this nonstationary gamma process has bounded mean value and standard deviation, its sample paths may still exceed $x_{lim}$. To fully characterize the reality, we need a stochastic process with every sample path that is bounded. In a rigorous probability language, this means that we need to ensure that the stochastic process is bounded not only *in the mean square sense*, but also *with probability* 1. The value of this study is to demonstrate the difference of two versions of gamma process and the effects of the difference in maintenance optimization.

## 2.2 Bounded transformed gamma process

Pulcini and his colleagues proposed a few versions of transformed gamma process with bounded sample paths (Fouladirad et al., 2023; Giorgio et al., 2015). Their approach of model construction is by translating a nonstationary gamma process to a bounded version through a simple transform function such as a shifted inverse function, a logarithm function or a tangent function. Here we offer an additional alternative.

Our approach of model construction is slightly different: We first employ a standard stationary gamma process as the kernel, and then use a Weibull function to translate the kernel process to the observable degradation process. Specifically, let $G(t)$ denote a standard gamma process with shape function $\alpha(t) = \alpha \cdot t$ and $\beta = 1$. Then the bounded transformed gamma process (BTGP), denoted as $X(t)$, is expressed as $X(t) = \mathcal{T}[G(t)]$, in which $\mathcal{T}[\cdot]$ represents a deterministic mapping expressed as:

$$\mathcal{T}[g_t] = x_{lim}\left(1 - \exp\left[-\left(\frac{g_t}{\theta_3}\right)^{\theta_2}\right]\right), \tag{4}$$

in which $g_t$ here represents a given sample path of the standard gamma process $G(t)$. Here, we reserve $\theta_1$ for the shape parameter $\alpha$ of the standard gamma process. Therefore, both the BT and the BN versions have the same number of model parameters, making model comparisons easier. Note that BTGP given by Equation (4) is a monotonically increasing process. When the transform function change to the following form:

$$\mathcal{T}[g_t] = x_{lim} \exp\left[-\left(\frac{g_t}{\theta_3}\right)^{\theta_2}\right], \tag{5}$$

the BTGP turns to a decreasing version: each sample path starts with $x_{lim}$ at time 0 and drops to 0 as time goes to infinity. Without stated otherwise, the increasing form is used in the following.

The Weibull-type function above is not chosen arbitrarily. It is built upon our extensive experience in deterioration modelling derived from many years of research and practice in infrastructure asset management and management system development for various infrastructure systems including bridges, buildings, pavements, water and wastewater systems, and nuclear power plants. Existing asset management systems use nonlinear regression deterioration models for continuous-state deterioration phenomena. Due to its inherent flexibility in adapting to various deterioration patterns, the Weibull function has been extensively used as the nonlinear regression model. However, such a nonlinear regression model has a key limit; that is, it does not account for the important temporal uncertainty as highlighted in Pandey et al. (2009). With a gamma process indexed time scale, the Weibull function allows to accurately describe both early-stage deterioration and long-term asymptotic behavior while maintaining interpretability of model parameters.

*2.2.1 Interpretations*

There are two ways to interpret the proposed BTGP model. The first one is by definition, i.e., the BTGP is constructed from a nonlinear transformation of a standard gamma process. This interpretation helps with Monte Carlo simulation. One can first generate random samples from the standard gamma process, and then apply the deterministic nonlinear transform to obtain the sample paths of the BTGP. Although $G(t)$ is unbounded, every sample path of $X(t)$ is bounded after the bounded nonlinear transformation given in Equation (4) or (5). Conversely, define an inverse transform as

$$\mathcal{G}[x_t] = \theta_3 \ln^{\frac{1}{\theta_2}}\left(\frac{x_{lim}}{x_{lim} - x_t}\right) \text{ or } \theta_3 \ln^{\frac{1}{\theta_2}}\left(\frac{x_{lim}}{x_t}\right). \qquad (6)$$

If $\mathcal{G}[X(t)]$ follows a standard gamma process, then $X(t)$ is a BTGP.

An alternative interpretation of the BTGP is given from the traditional regression modelling viewpoint. Note that several pavement and bridge asset management software packages employ a nonlinear regression model in exactly the same form as given by Equation (4), with the left-hand side replaced by the degradation and the $g_t$ term replaced by service age $t$. Therefore, the BTGP can be considered as a nonlinear regression model with a random 'internal aging clock', which follows the standard gamma process $G(t)$. Compared with the traditional regression model, the BTGP includes an explicit mechanism to characterize the temporal uncertainty in performance deterioration of infrastructure systems. Note that the notion of internal or virtual aging clock is not new in the literature of accelerated degradation modelling; see (Lawless et al., 1995; Lee & Whitmore, 1993; Whitmore & Schenkelberg, 1997) for examples.

*2.2.2 Distributions of the bounded transformed gamma process*

The marginal distribution of $X(t)$ can be readily derived as follows:

$$f_{X(t)}(x) = \mathcal{G}'(x) \cdot f_{G(t)}(\mathcal{G}(x); \alpha t, 1)$$

$$= \frac{\theta_3}{\theta_2(x_{lim} - x)} \ln^{\frac{1}{\theta_2}-1}\left(\frac{x_{lim}}{x_{lim} - x}\right) \cdot f_{G(t)}\left(\theta_3 \ln^{\frac{1}{\theta_2}}\left(\frac{x_{lim}}{x_{lim} - x}\right); \alpha t, 1\right), \tag{7}$$

for $0 \leq x \leq x_{lim}$. In Equation (7), $f_{G(t)}(\cdot; \alpha t, 1)$ represents the probability density function of a gamma distributed random variable with shape $\alpha t$ and scale 1. Apparently, there is no explicit expression for the mean value or standard deviation, but they can be evaluated numerically.

To study the distribution of increments, let us define $\Delta X(t) = X(t + \Delta t) - X(t)$ and $\mathcal{G}(\Delta x|x_t) = \mathcal{G}(x_t + \Delta x) - \mathcal{G}(x_t)$. With a given $x_t$, $d\mathcal{G}(\Delta x|x_t)/d\Delta x = \mathcal{G}'(x_t + \Delta x)$. Then, conditional on $X(t) = x_t$, the probability density function of the increment can be similarly derived as

$$f_{\Delta X(t)}(\Delta x|x_t) = |\mathcal{G}'(x_t + \Delta x)| f_{G(t)}(\mathcal{G}(\Delta x|x_t); \alpha \Delta t, 1), \tag{8}$$

for $0 < \Delta x < x_{lim} - x_t$. This expression is useful for constructing the likelihood function for the BTGP modelling, which is discussed in the next section. Since the conditional distribution of the increment $\Delta X(t)$ is dependent on $X(t)$, increments of the BTGP over disjoint time intervals are not independent. Nevertheless, the BTGP is still Markovian as Equation (7) shows.

Suppose that an infrastructure system degrades in accordance with the BTGP $X(t)$. The system is said to fail whenever $X(t)$ exceeds a predefined threshold $\xi$. Then the first-passage failure time, $T \stackrel{\text{def}}{=} \inf_{t} \{X(t) \geq \xi\}$, is a random variable that has the following survival function:

$$S(t) = \Pr(X(t) < \xi) = F_{X(t)}(\xi) = GA(\mathcal{G}(\xi); \alpha t, 1), \tag{9}$$

where $GA(\cdot; \alpha, 1)$ represents the cumulative distribution function (CDF) of a gamma random variable with shape $\alpha t$ and scale 1.

*2.2.3 Likelihood function*

As mentioned earlier, the BTGP deterioration model has three basic parameters: $\alpha = \theta_1$ for the standard gamma process, and $\theta_2$ and $\theta_3$ for the nonlinear transform function. Denote $\boldsymbol{\theta} = (\theta_1, \theta_2, \theta_3)$. Maximum likelihood method is a common and effective statistical inference technique for parameter estimation. The construction of the likelihood function is the key. For an infrastructure system of $m$ statistically independent units, each of $n_i$ inspections, from which the infrastructure condition at inspection $t_{ij}$ is assessed and denoted as $x_{ij}$ ($i = 1, 2, \ldots, m; j = 1, \ldots, n_i$), the likelihood function of the BTGP model is expressed as

$$L(\theta \mid Data) = \prod_i \prod_j L(\theta \mid \Delta x_{ij}) = \prod_i \prod_j f_{\Delta X_i(t_j)}(\Delta x_{ij} \mid x_{i,j-1}) \qquad (10)$$

where $\Delta x_{ij} = x_{ij} - x_{i,j-1}$ with $x_{i,0} = 0$ for all $i$, and $f_{\Delta X_i(t_j)}(\Delta x_{ij} \mid x_{i,j-1})$ is the probability density function of increments as defined in Equation (7). For the BTGP model, the $x_{lim}$ is given.

## *2.3 Qualitative comparisons*

In addition to the BNGP and the proposed BTGP model, there are six additional BTGP options proposed by Giorgio and Pulcini (2024). For the ease of reference, all eight candidate models are listed in Table 1, where BTGP denotes our proposed model, whereas BTGP1 to BTGP6 are the six variants corresponding to $g_1(w)$ to $g_6(w)$ in Giorgio and Pulcini (2024). Note that the $g(w)$ functions here can be understood as the inverse of transformation function $\mathcal{T}$ in this study. Note particularly that all BTGP models in Giorgio and Pulcini (2024) in their original forms have a nonstationary underlying gamma process with a scale parameter $\beta$. For consistency, this study takes $\beta = 1$ because otherwise the scaling effects would be cancelled out by the scale parameter in the (state-) transform function. This study does not consider the transformed gamma process (TGP) used in (Giorgio & Pulcini, 2024) for it has an unbounded mean curve.

**Table 1 Candidates Models**

| Model | Transform function | Underlying Gamma Process | |
|---|---|---|---|
| | | Shape function $\alpha(t)$ | Scale parameter $\beta$ |
| BNGP | -- | $\frac{x_{lim}}{\theta_1}\left(1 - \exp\left[-\left(\frac{g_t}{\theta_3}\right)^{\theta_2}\right]\right)$ | $\theta_1$ |
| BTGP | $\mathcal{T}_0[g_t] = x_{lim}\left(1 - \exp\left[-\left(\frac{g_t}{\theta_3}\right)^{\theta_2}\right]\right)$ | $\theta_1 t$ | 1 |
| BTGP1 | $\mathcal{T}_1[g_t] = x_{lim}\left[1 - \exp\left(-\frac{g_t}{\theta_3}\right)\right]$ | $\left(\frac{t}{\theta_1}\right)^{\theta_2}$ | 1 |
| BTGP2 | $\mathcal{T}_2[g_t] = x_{lim}\left(\frac{g_t/\theta_3}{1+g_t/\theta_3}\right)$ | $\left(\frac{t}{\theta_1}\right)^{\theta_2}$ | 1 |
| BTGP3 | $\mathcal{T}_3[g_t] = x_{lim}\left[\frac{2}{\pi}\tan^{-1}\frac{g_t}{\theta_3}\right]$ | $\left(\frac{t}{\theta_1}\right)^{\theta_2}$ | 1 |
| BTGP4 | $\mathcal{T}_4[g_t] = x_{lim}\left[1 - \exp\left(-\frac{g_t}{\theta_3}\right)\right]^{\theta_4}$ | $\left(\frac{t}{\theta_1}\right)^{\theta_2}$ | 1 |
| BTGP5 | $\mathcal{T}_5[g_t] = x_{lim}\left(\frac{(g_t/\theta_3)^{\theta_4}}{1+(g_t/\theta_3)^{\theta_4}}\right)$ | $\left(\frac{t}{\theta_1}\right)^{\theta_2}$ | 1 |
| BTGP6 | $\mathcal{T}_6[g_t] = x_{lim}\left[\frac{2}{\pi}\tan^{-1}\frac{g_t}{\theta_3}\right]^{\theta_4}$ | $\left(\frac{t}{\theta_1}\right)^{\theta_2}$ | 1 |

*2.3.1 Comparison among bounded transformed gamma process models*

As presented in Table 1, the seven BTGPs share the same model structures in the sense that they all have an underlying gamma process and a state-transform function. Therefore, all BTGPs belong to the same category. The major difference lies in parameterization. Our proposed BTGP uses a stationary gamma process, whereas the nonlinearity of the mean curve is characterized by the Weibull-type transform function. In contrast, the other six BTGPs use a nonstationary gamma process and different transform functions with various complexities. Note that the BTGP4~6 models are a further parametrized variant of BTGP1~3, respectively.

From the functional form it is difficult to tell which BTGP model is more flexible than the other in fitting actual degradation data. We defer this to Section 3 for empirical studies. It is worth highlighting one important feature of the proposed BTGP; that is the stationary underlying gamma process. This feature suggests that the proposed model is, following the terminology of Guida and Pulcini (2011), age-independent and state-dependent. The age-independent feature makes the model easy to use in infrastructure asset management, because when maintenance and rehabilitation activities are implemented, asset performance may be only partially improved, and therefore, the dilemma of tracing the effective age after maintenance and rehabilitation does not exist. For detailed explanation of the dilemma, interested readers are referenced to Zhang and Yuan (2021). For the same reason, the age-independent feature also makes the computation of the likelihood function and state prediction easier than an age-dependent model.

*2.3.2 Comparison between BNGP and the proposed BTGP*

While the BNGP and the proposed BTGP models share several common features, each has its own salient features. Table 2 compares qualitatively the major features of the two models. Both have continuous and monotonically increasing sample paths. In addition, both models have a bounded mean function in the sense that the mean value converges to a constant as time goes to infinity. The specific expressions of the distribution functions of the state variable $X(t)$, the increments $\Delta X(t)$, and the first-passage lifetime $T$ are also listed for easy reference.

There are two major differences. Firstly, the sample paths of the BTGP are bounded, whereas those of the BNGP are not. Secondly, the BNGP has a monotonically increasing and bounded variance function. However, the variance of the BTGP is no longer a monotonic function of time. Rather, the variance increases first, and then decreases and converges to zero as time goes to infinity. This is natural because as time goes by all sample paths converge to the same upper limit, rendering the variance to zero.

**Table 2: Summary and comparison of major features of the BNGP and BTGP**

|  | **BNGP** | **BTGP** |
|---|---|---|
| **Full name** | Bounded nonstationary gamma process | Bounded transformed gamma process |
| **Temporal dependence** | Strong Markovian with independent increments | Markovian; increments are not independent |
| **Sample path** | Continuous; monotonically increasing; theoretically unbounded | Continuous; monotonically increasing; bounded |
| **Mean function** | Monotonically increasing & bounded; analytical expressed | Monotonically increasing & bounded; can only be numerically evaluated |
| **Variance function** | Monotonically increasing & bounded | Nonmonotonic & bounded - increasing and then decreasing to zero; can only be numerically evaluated |
| **PDF of $X(t)$** | Equation (1) | Equation (7) |
| **PDF of $\Delta X(t)$** | Equation (1) with $\alpha(t)$ replaced by $\Delta\alpha(t)$; dependent on $t$ | Equation (8); conditional on $X(t)$ |
| **Lifetime distribution** | $S(t) = GA(\xi; \alpha(t), \beta)$ | Equation (9) |

To graphically demonstrate the differences, an arbitrary BTGP model is preselected at first, and then a least square method is used to determine the parameters of the BNGP so that the mean curves of the two models are matched. Figure 1(a) and (b) show three random sample paths and the mean curve (dashed line) of the bridge condition index (BCI) of the BNGP and BTGP models, respectively. As one can see, the two mean curves are very close (note that the y-axis ranges of the two graphs are different). However, all three sample paths of the BTGP model converge strictly to 0, whereas the sample paths of the BNGP model may drop below zero.

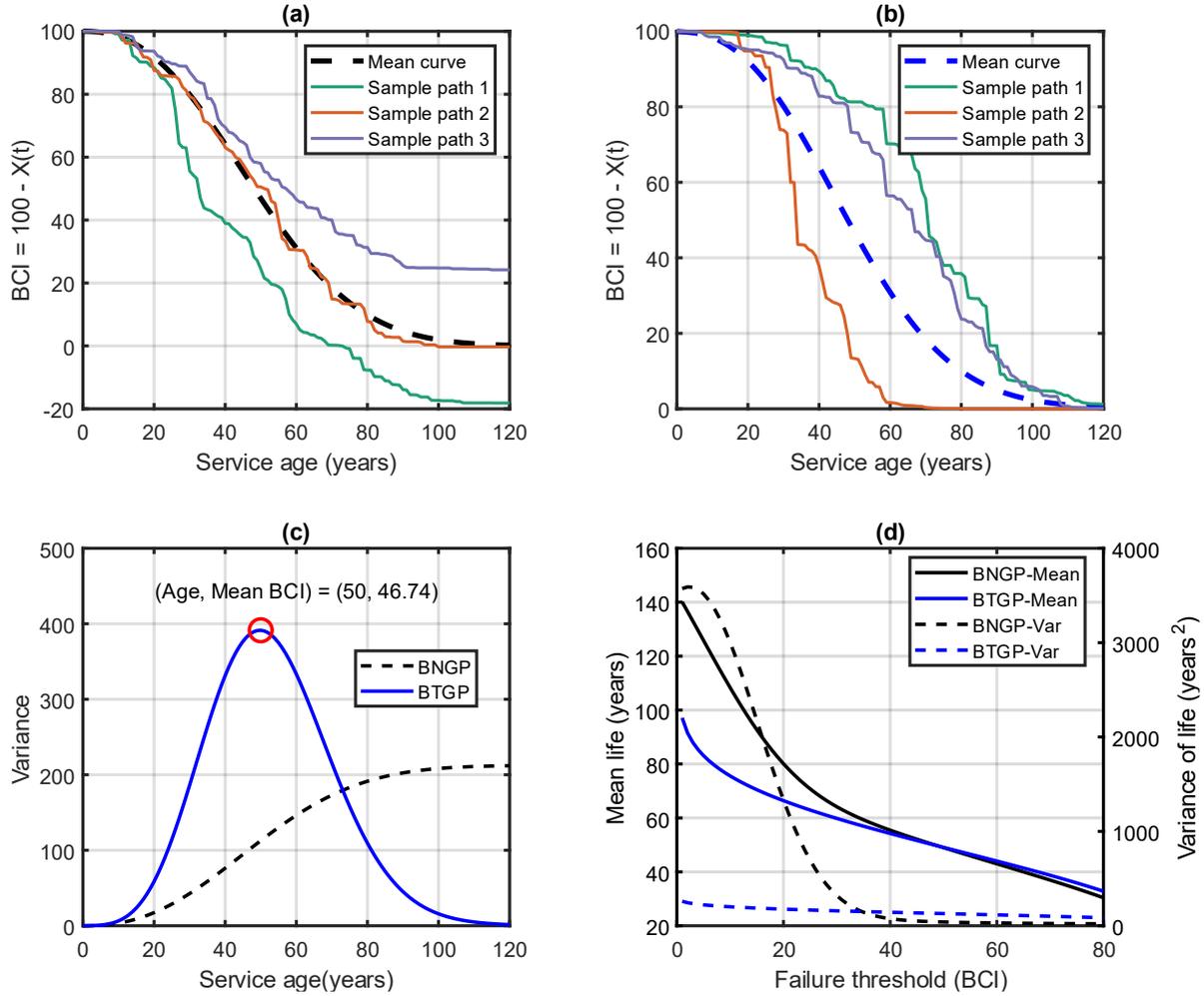

**Figure 1 Comparisons of samples paths, variance functions, and average first-passage time between the BNGP and BTGP with a same mean curve.**

The variance functions of the two models are compared in Figure 1(c). The variance function of the BNGP is monotonically increasing, whereas that of the BTGP is non-monotonic. Since both models can be considered as being benchmarked against the same dataset, the upper bound of the BNGP variance is lower than the maximum variance of the BTGP.

Because of these two major differences, the first-passage lifetime distributions derived from the two models also show various degrees of difference. To illustrate this difference, the mean and variance of lives at different failure threshold $\xi$ values of the two models are computed and plotted in Figure 1(d). Unsurprisingly, the difference of mean and variance of lives get more significant as the failure threshold $\xi$ gets lower. This is because when the failure threshold $\xi$ is smaller, the difference of the two models in the variance at the corresponding mean condition state increases as well. It should be noted that Figure 1 (d) just serves as an illustrative example, the magnitude of the difference may vary depending on the model parameters.

Note that the preceding comparison is based on the condition that the two models' mean curves are matched. One may speculate whether both the mean curve and the variance function of the BTGP model can be matched with the corresponding curves of the BNGP counterparts (e.g., through the use of some sort of method of moments). After all, the BNGP is conceptually simpler than the BTGP. One may immediately notice that a full match of the variance over the whole range is logically impossible, because, as pointed out earlier, the BTGP has a non-monotonic variance function whereas the BNGP's variance function is always non-decreasing. Nevertheless, one may further speculate that if the ascending segment of the BTGP variance function is long enough to effectively cover the surviving state of the system before the failure threshold. Then, maybe the BNGP model can effectively approximate the BTGP in the surviving range of the system. This seems possible, because in most cases of equipment, pipelines, transportation, and water and wastewater infrastructure systems with bounded phenomena, the system performance failure is often defined when the critical performance measure deteriorates to 40%~70% of the total range. This brings an interesting question to investigate: at what level of the mean value does the variance or standard deviation reach its maximum value? This is the topic of the next subsection. The mean value (or $\mu$) at the time when the variance reaches its peak (i.e., maximum variance) is abbreviated as MUMV, read as "mu at MV".

*2.4 Mean value at maximum variance of the BTGP*

The MUMV is a critical parameter when comparing the BNGP with the BTGP. This is because if the BNGP's variance function can match the ascending part to the BTGP's variance function, and moreover, if the ascending part takes the whole effective asset life, then one may not need the BTGP model and can still choose to use the relatively simpler BNGP model. Therefore, this subsection is dedicated to the MUMV regarding its sensitivity to the BTGP model parameters.

The MUMV is determined numerically as follows. First, the time when the variance reaches the maximum value is determined. Then, the mean value of the deterioration at that time is calculated. Figure 2 shows the plots of MUMV against the three model parameters ($\theta_1 = \alpha, \theta_2$ and $\theta_3$ individually, with the other two parameters being in the baseline value: $\theta_1 = 0.38$, $\theta_2 = 3.22$ and $\theta_3 = 21.18$. The results indicate that the MUMV varies very little with $\theta_1$ and $\theta_3$, but significantly with $\theta_2$. A detailed analytical investigation shows that the MUMV is in fact independent of $\theta_1$ (i.e., $\alpha$). The little variation shown in Figure 2 (a) is caused by numerical errors. Using the following definitions:

$$\mathrm{E}[X(t)] = \mathrm{E}_{G(t)}\{\mathcal{T}[G(t)]\}, \qquad \mathrm{E}[X^2(t)] = \mathrm{E}_{G(t)}\{\mathcal{T}^2[G(t)]\} \qquad (11)$$

one can readily show that the mean value function and the variance function both are dependent of $\tau = \alpha t$. In other words, at a given $\tau$ value, the mean and variance will not change for different $\alpha$ values. For a given set of $\theta_2$ and $\theta_3$, the maximum variance will stay constant, although the time of the maximum will change with $\theta_1$ or $\alpha$. This once again confirms one of our earlier interpretations that the underlying gamma process plays a role of random time scaling of the deterioration process.

The influence of $\theta_2$ on the MUMV is not surprising because it is the shape parameter of the transform function. However, the influence of $\theta_3$ is not so intuitive. Our careful examination confirms that the influence of $\theta_3$ on MUMV is not caused by numerical fluctuation. Note that the underlying gamma process $G(t)$ has both mean and variance functions equal $\alpha t$. Suppose that the mean and variance functions of $X(t)$ are denoted as $\mu(\tau;\theta_2,\theta_3)$ and $\sigma^2(\tau;\theta_2,\theta_3)$. Due to Jensen's inequality, $\mu(\tau;\theta_2,\theta_3) \neq \theta_3\mu(\tau;\theta_2,1)$ and $\sigma^2(\tau;\theta_2,\theta_3) \neq \theta_3^2\sigma^2(\tau;\theta_2,1)$.

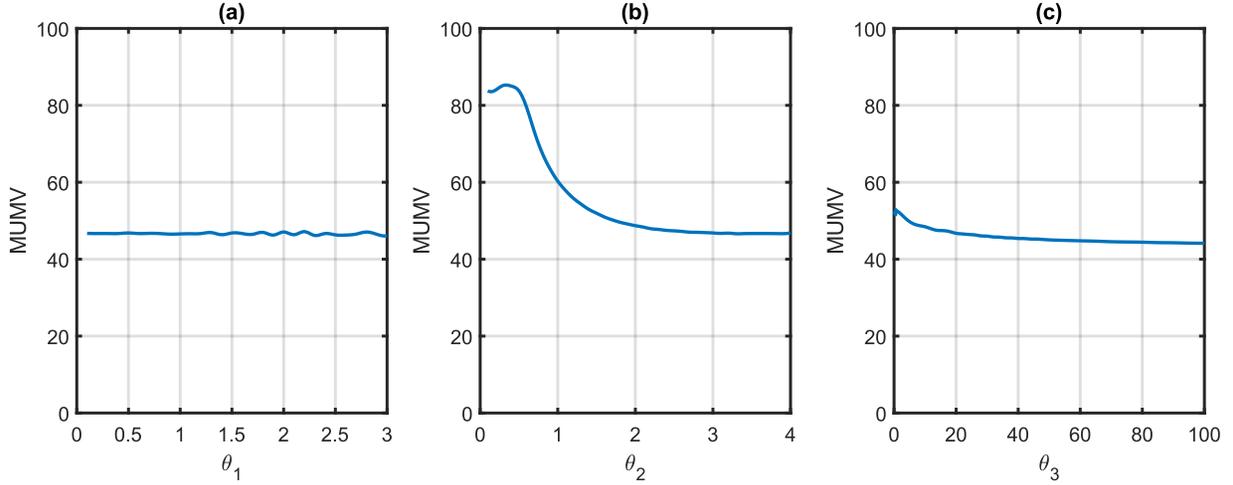

**Figure 2 MUMV against $\theta_1$, $\theta_2$ and $\theta_3$ (Baseline: $\theta_1 = 0.38$, $\theta_2 = 3.22$ and $\theta_3 = 21.18$)**

To further show the variability of MUMV, a more thorough two-dimensional sensitivity analysis was carried out with $\theta_2$ ranging from 0.1 to 4 and $\theta_3$ from 1 to 100. The results are presented as a contour plot shown in Figure 3. The plot shows that within a wide range of model parameters, the MUMV ranges from 45 to 90. Note that the upper limit $x_{lim}$ is taken as 100 and a decreasing version of the BTGP is simulated here. This range means that the increasing portion of the variance function of the BTGP will terminate when the mean deterioration takes 10 to 55 percent of the entire range of the deterioration. After that, the variance function starts to decrease with time.

To our best knowledge, that the MUMV of a decreasing (increasing) BTGP has a lower (upper) bound is a phenomenon that has not been reported elsewhere. This suggests that the MUMV is a critical parameter that reflects the inherent flexibility of the bounded transformed gamma process model. For example, suppose the estimated empirical variance function peaks at a mean value outside the inherent MUMV range of a given BTGP model (as shown in Figure 3), then this BTGP model, no matter what values its parameters take, will not be able to adequately capture the observed second-order variability. How to expand the range of the MUMV seems to deserve further investigation. Different transform functions may change the range, but we leave it for future study.

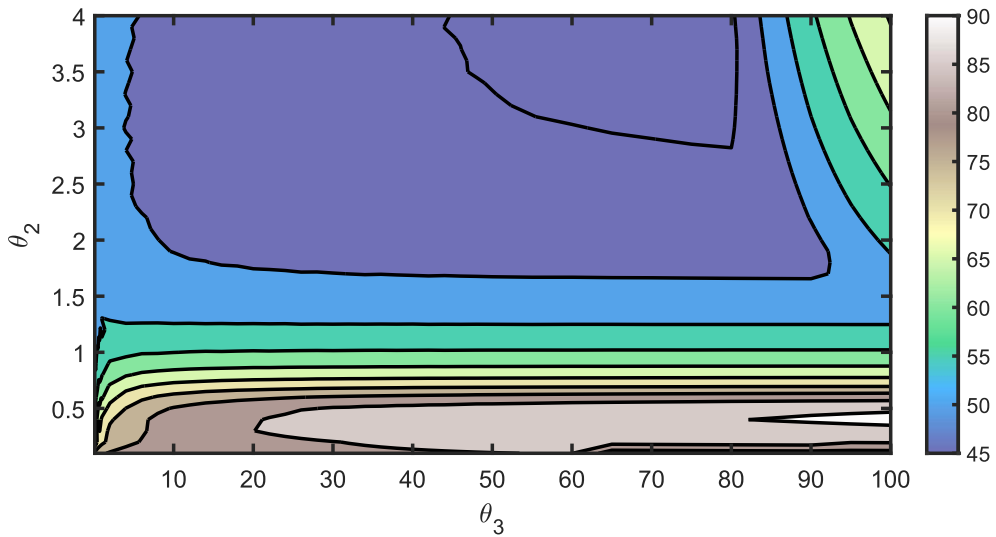

**Figure 3 MUMV under various combination scenarios of $\theta_2$, $\theta_3$ under a given $\theta_1 = 0.38$**

## 3 Empirical study

The empirical study uses a dataset of the Ontario Open BCI data taken from the Province of Ontario's *Open Government Portal- Bridge conditions*. BCI stands for Bridge Condition Index, which is a bridge condition measure ranging from 100 (best condition) to 0 (worst). The Open BCI data contain the historical BCI data of all 2,994 bridges in the provincial highway network of Ontario, Canada covering the bridge service age ranging from 0 to 80 years. Due to disparate maintenance and rehabilitation activities, the BCI profiles are not necessarily monotonic. Data cleansing was carried out to identify the natural deterioration path with a monotonic decline trend.

Measurement errors were neglected in this process.

There are two ways of using the database. One is to develop a parameterized population model for all bridges in the inventory using all data, and the other is to develop individual, bridge-specific models. Since the focus of the study is placed on the flexibility of the candidate models, we take the second approach. For that purpose, only those deterioration paths with six or more records were kept so that a reliable bridge-specific model can be estimated. This yielded 267 bridges (each bridge having 6 to 10 data points covering 12 to 20 years), whose BCI deterioration paths are depicted in Figure 4.

To determine the most flexible model, each BCI profile is fitted separately using the eight candidate models, and the model with the minimum AIC value was chosen as the best model for the bridge. After that, count the total cases of a candidate model being selected as the best model, and divide the count by 267 to yield the percentage of the best model for the candidate model. The distribution of the percentages of the best model is shown in the bar chart in Figure 5. Among the eight candidates, the proposed BTGP model takes 42% of the cases. This is eight times more than the frequency of the BNGP model. The result also corroborates the early finding made by Giorgio and Pulcini (2024) that BTGP2 outperform BTGP4 and BTGP 6 models. Meanwhile, although the BTGP2 is better than the BNGP model, it is much worse than the proposed BTGP model in terms of the percentage of best model.

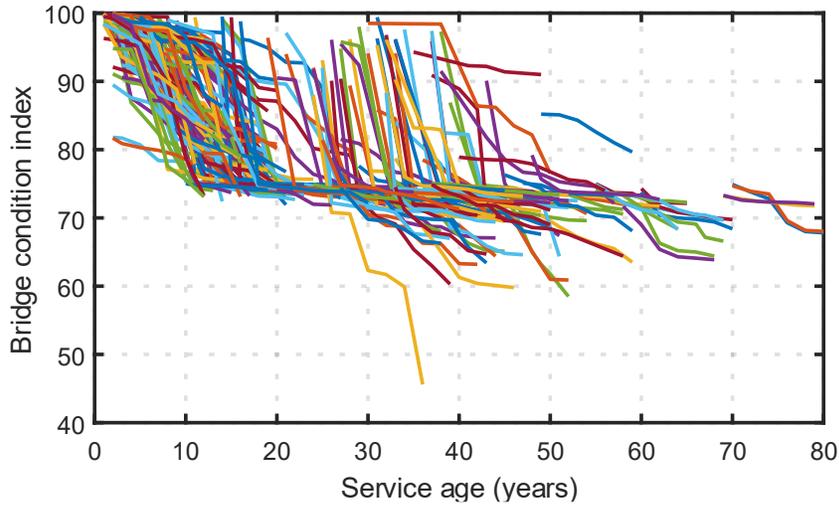

**Figure 4 Observed BCI deterioration paths of 267 bridges (Observations that pertain to the same bridge are connected for graphical convenience)**

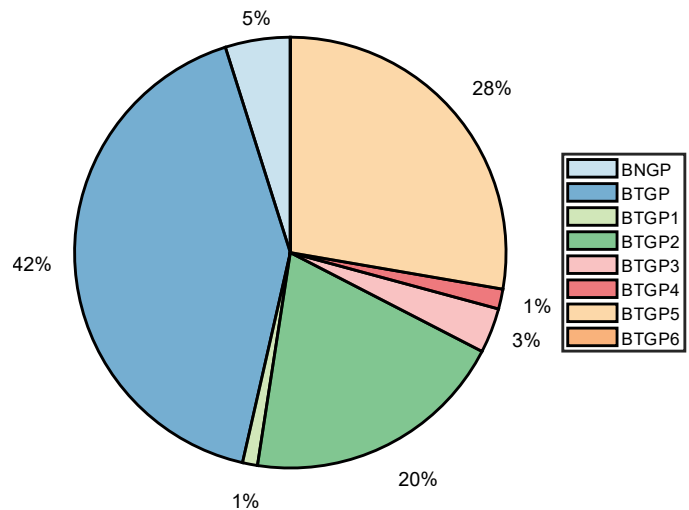

**Figure 5 Distribution of the candidate models being selected as the best model**

Note that the proposed BTGP and BTGP4 models equal only when $\theta_2 = \theta_4 = 1$. Since $\theta_4$ is applied outside of the square bracket, it does not provide the capability of altering the overall shape of the degradation pattern. With $\theta_2$ applied inside the exponent term, our proposed model allows for various patterns. This explains why our model has smaller AIC values than BTGP4 in most cases that we tested.

# 4 Impact to asset management decision-making

There was a saying, often accredited to George Box (a famous statistician), stating that 'all models are wrong, but some useful.' Aligning with this utility spirit, Field and Grigoriu (2007) suggest that decision-theoretical criteria, rather than information criteria, be used for model selection. This section further compares the BNGP and the proposed BTGP model from the decision-making perspective. Specifically, we want to examine how significant the difference in the asset management decisions can be when different deterioration models that are calibrated by the same set of empirical data are applied to age- and condition-based replacement policies. Note that the performance of the other six BTGP models discussed precedingly have been studied in (Giorgio & Pulcini, 2024) and thus are not included here, for it is not our intention to re-evaluate those models. In addition, asset management practices may include many other maintenance and rehabilitation policies other than replacement (or reconstruction) policies. The purpose of this section is to illustrate the impact of two competing models on decision making using replacement policies as an example. Comprehensive investigations are left for future studies.

## *4.1 The data and deterioration models*

One of the bridges with natural degradation path from the preceding empirical study is taken as a case study. The bridge was constructed in 1997, and a total of 8 inspections from 2000 to 2020 were carried out, resulting in 8 BCI records. The BCI deterioration data are then used to fit the traditional BNGP and the proposed BTGP model, with $x_{lim}$ set to 100. Figure 6 illustrates the observed data, the mean and variance of the two estimated models, as well as the 95% predictive bands. The two mean curves are very close; however, the variance curves match very well for the first 5 years only and after that they diverge and differ drastically. As a result, the 95% predictive band of the BTGP model widens up for the first 30 years and then slowly narrows down afterwards.

In contrast, the predictive band of the BNGP keeps widening. If the band continues beyond the 100 years of service age, the lower bound will drop below zero, which is apparently unrealistic.

Based on the last inspection result of BCI = 79.68 at $t = 15$ years, the probability distribution of the BCI decrements in the remaining 85 years and the distribution of the remaining service life with a failure threshold $\xi = 40$ are calculated using the two estimated models and shown in Figure 7. The difference in the two BCI distributions at the left panel of Figure 7 is explained by the similar mean values and significantly different variances shown in Figure 6. Giorgio and Pulcini (2024) did a similar comparison between a nonstationary gamma process and one of the BTGP models they proposed, and the difference of the two distributions was much greater, which was because their nonstationary gamma process model has an unbounded mean function.

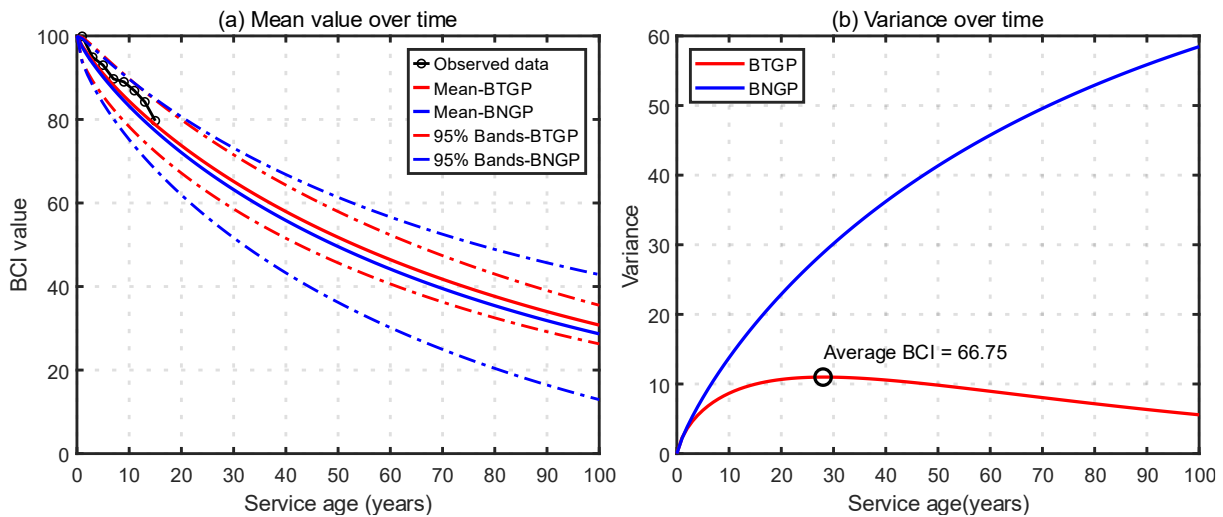

**Figure 6 Mean and variance under the BNGP and BTGP models**

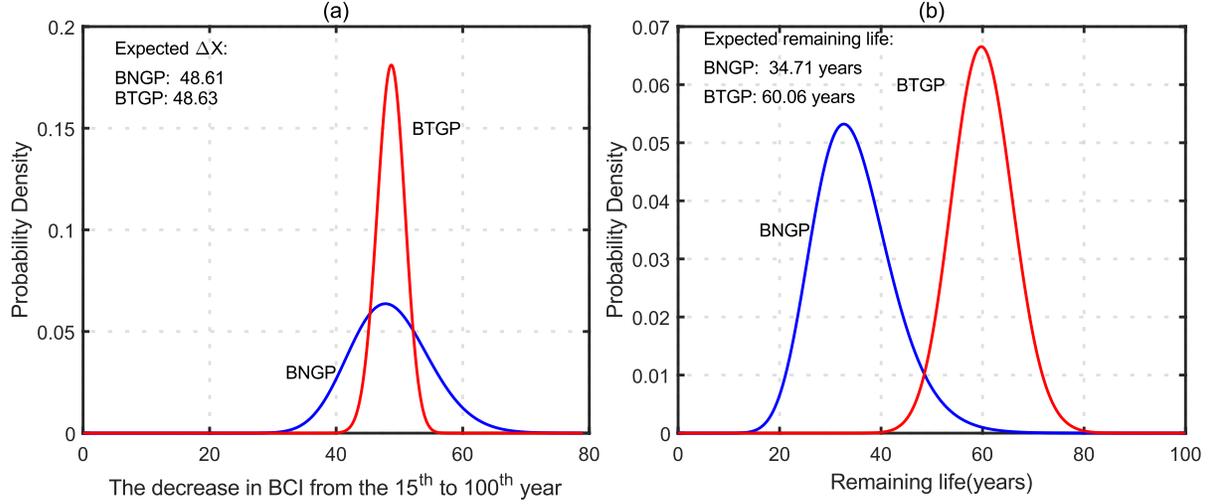

**Figure 7 Comparisons of ΔX and remaining lifetime distributions**

The remaining service life of the bridge shows greater discrepancy between the two models: the BNGP model predicts a much smaller mean than the BTGP model does, whereas the trend for variability is opposite. This difference has a direct impact on the age- and condition-based policies, which are discussed next.

*4.2 The age-based replacement policy*

The age-based replacement (ABR) policy is one of the simplest strategies for deteriorating systems. Under this strategy, a system is replaced when it reaches a fixed service age or fails, whichever comes first. The periodic replacement age is denoted as $t_R$. In addition, the preventive replacement (PR) cost is denoted as $C_R$, and the failure replacement cost as $C_F$, which is generally much greater than $C_R$. The primary objective of the ABR strategy is to determine the optimal replacement age $t_R$ so as to minimize the mean lifecycle cost (LCC) rate over an infinite planning horizon. According to renewal theory, the mean LCC rate $K(t_R)$ can be expressed as:

$$K(t_R) = \lim_{t \to \infty} \frac{\mathrm{E}[K(t)]}{t} = \frac{C_R \int_0^{t_R} S(t)\mathrm{d}t + C_F \int_0^{t_R}[1 - S(t)]\mathrm{d}t}{\int_0^{t_R} S(t)\mathrm{d}t} \qquad (12)$$

where $K(t)$ represents the total cost from time 0 to $t$, and $S(t)$ is the survival function.

With $C_R=$ \$100 and $C_F =$ \$500, Figure 8 compares the mean LCC rates at different periodic replacement ages between the BNGP and BTGP models. The optimal ages and their corresponding minimum LCC rates are identified as well. The optimal replacement ages for BNGP and BTGP are 45 years and 59 years, respectively, with BNGP exhibiting a higher average LCC rate and much lower replacement age. The reason for this is that the BNGP model involves higher uncertainty than the BTGP model, as discussed above.

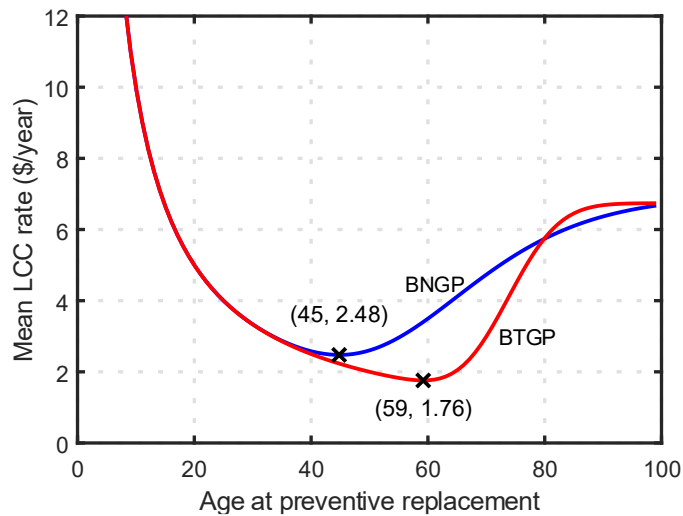

**Figure 8  Comparison of the mean LCC rate curves for the ABR policy**

Since the failure threshold affects the failure rates of the system lifetime, the sensitivity of the optimal replacement age and the minimum average LCC rate to the failure threshold is studied and the patterns are shown in Figure 9. As the failure threshold increases, the optimal replacement ages for both models decrease and the gap also reduces. This is because as the failure threshold moves up, the difference between the two deterioration models reduces. Nevertheless, the mean LCC rate derived from the BNGP model remains consistently higher than that from the BTGP model because the BNGP model results in earlier replacement.

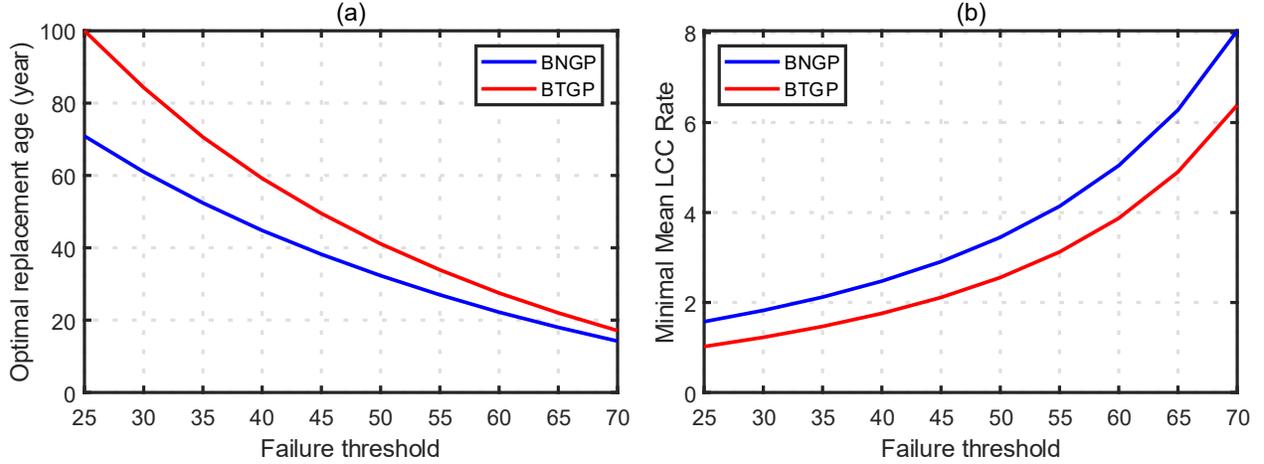

**Figure 9 Optimal replacement ages and LCC rates under different failure thresholds**

### *4.3 The condition-based replacement policy*

The condition-based replacement (CBR) strategy involves periodic inspections at a fixed time interval $t_I$. Depending on the inspection results, the system may face one of the three options: do nothing, preventive replacement, and failure replacement. If the performance is found above a preventive maintenance threshold $\xi_R$, then do nothing. If the system is found to have failed, i.e., the performance is below the preset failure threshold $\xi$, then a failure replacement must be done. If the performance is found in between $\xi$ and $\xi_R$, then a preventive replacement is performed. The objective of CBR strategy optimization is to find an optimal inspection interval $t_I$ and replacement thresholds $\xi_R$ together that minimizes mean LCC rate, resulting in a two-dimensional optimization problem. According to renewal theory, the mean LCC rate $K(t_I)$ can be expressed as:

$$K(t_I) = \lim_{t \to \infty} \frac{\mathrm{E}[K(t)]}{t} = \frac{\sum_{n=1}^{+\infty}\left[(C_R + nC_I)p_R(n) + [C_F + (n-1)C_I]p_f(n)\right]}{\int_0^{t_R} S(t)\mathrm{d}t} \tag{13}$$

where $p_R(n)$ denotes the probability of preventive replacement at $nt_I$, and $p_f(n)$ denotes the probability of corrective replacement between $[(n-1)t_I, nt_I]$. This study follows the detailed procedure developed in Pandey et al. (2009), with necessary modifications for the BTGP model.

*4.3.1 Baseline results*

The baseline study takes $C_I = \$1$, $C_R = \$100$ and $C_F = \$500$. Figure 10 shows the results of the two-dimensional optimization with a baseline failure threshold of $\xi = 40$. The contour plot indicates that the optimal point for the BNGP model is $t_I^{BNGP} = 6.3$ and $\xi_R^{BNGP} = 47$, with the minimum mean LCC rate of $1.91/yr, whereas for the BTGP model, $t_I^{BTGP} = 8.5$, $\xi_R^{BTGP} = 46$, and minimum LCC rate $1.68/yr. Although the optimized preventive maintenance thresholds are very close, the BTGP model suggests a smaller inspection frequency, primarily due to the smaller variability embedded in the model than the BNGP counterpart.

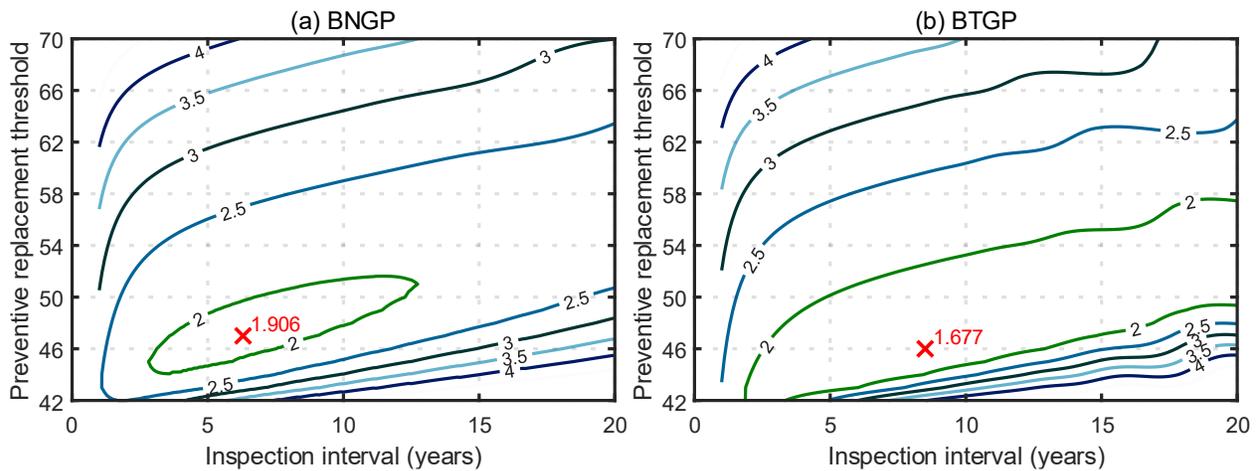

**Figure 10 Contour with the minimal LCC rate of BNGP and BTGP models**

*4.3.2 Effect of failure thresholds*

Similar to the ABR strategy, the sensitivity of the CBR strategy to the failure threshold $\xi$ is also studied and the results are shown in Figure 11. As expected, as the failure threshold decreases, the difference between the BNGP and BTGP models in terms of the optimal inspection interval, preventive replacement threshold, and LCC rate becomes more significant. When the failure threshold is below 40, the optimal inspection interval for the BNGP model hovers between 4 and 6 years, while the BTGP model allows a much longer inspection interval that is greater than 10 years. Additionally, as the failure threshold decreases, the optimal PR threshold for the BTGP model increasingly approaches the failure threshold, and the difference from the BNGP model becomes more pronounced. As a result, the difference in terms of the minimum mean LCC rate is even more significant when the failure threshold is below 40.

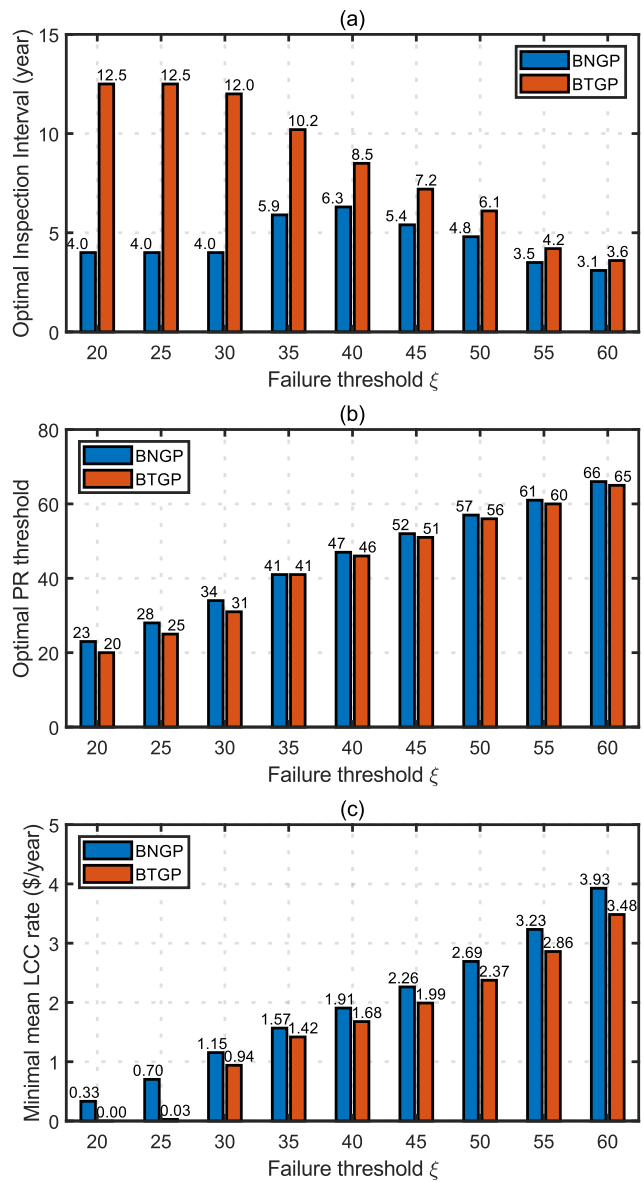

**Figure 11 LCC rates and inspection intervals under different failure thresholds**

*4.3.3 Effect of degradation paths*

The preceding analysis was based on only one single bridge. To enhance the study, two additional bridges of different degradation patterns were selected from the same database used in Section 5. The objective is to examine how the different degradation patterns would alter the major comparative findings about the two deterioration models. Figure 12 depicts the observation data, mean curves and 95% predictive bands. The modelling results for these two bridges as well as the previous, baseline bridge are shown in Table 3. It is worth highlighting the shape parameter $\theta_2$ of the BTGP models for the three bridges: 0.84, 1.30, and 2.81, respectively. With all decision-making parameters remaining unchanged, the optimal ABR and CBR results are summarized in Table 4. Three major observations are highlighted below:

**Table 3 Modelling results for two bridges**

| No. | BNGP model | | | BTGP model | | |
|---|---|---|---|---|---|---|
| | $\theta_1$ | $\theta_2$ | $\theta_3$ | $\theta_1$ | $\theta_2$ | $\theta_3$ |
| Baseline | 0.82 | 0.83 | 76.51 | 1.66 | 0.84 | 136.12 |
| Bridge #1 | 2.63 | 1.28 | 33.75 | 0.81 | 1.30 | 26.84 |
| Bridge #2 | 1.78 | 2.37 | 32.60 | 1.30 | 2.81 | 41.16 |

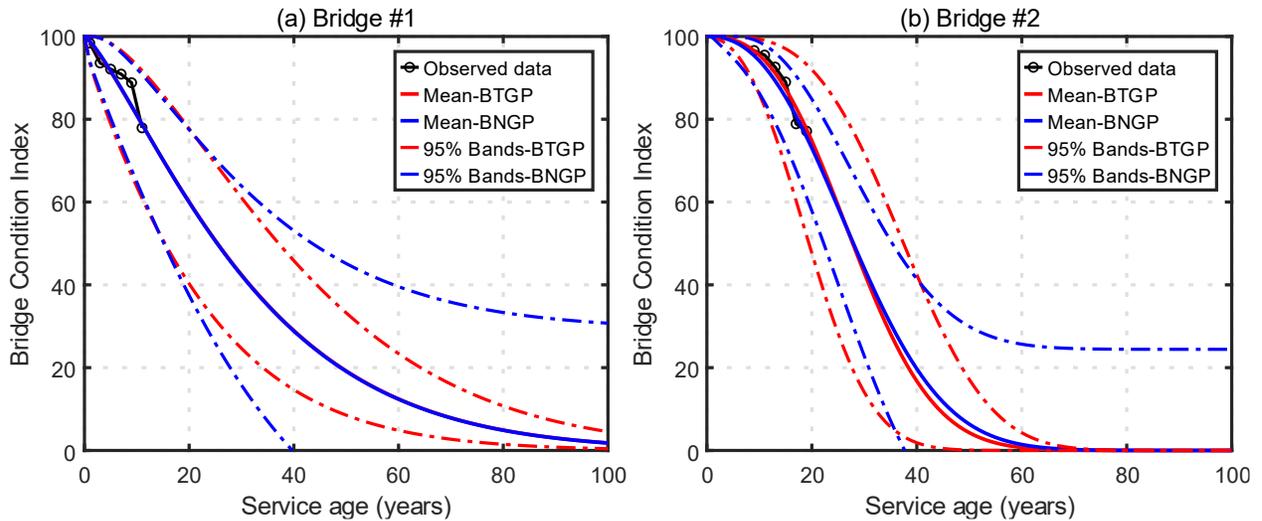

**Figure 12 BNGP and BTGP modelling results for two real degradation paths**

**Table 4 ABR-based and CBR-based inspection and replacement results for two bridges**

| No. | BNGP model | | | | | | BTGP model | | | | | |
|---|---|---|---|---|---|---|---|---|---|---|---|---|
| | RL (yr) | $t_R$ (yr) | $K_{opt}^{ABR}$ ($/yr) | $t_I$ (yr) | $\xi_R$ | $K_{opt}^{CBR}$ ($/yr) | RL (yr) | $t_R$ (yr) | $K_{opt}^{ABR}$ ($/yr) | $t_I$ (yr) | $\xi_R$ | $K_{opt}^{CBR}$ ($/yr) |
| Baseline | 34.7 | 44.8 | 2.48 | 6.3 | 47 | 1.91 | 60.1 | 59.2 | 1.76 | 8.5 | 46 | 1.68 |
| Bridge #1 | 10.06 | 19.8 | 5.76 | 2.7 | 53 | 4.47 | 8.97 | 20.8 | 5.43 | 2.5 | 52 | 4.46 |
| Bridge #2 | 16.27 | 23.4 | 4.32 | 6 | 60 | 4.79 | 5.64 | 21.7 | 5.01 | 2.6 | 60 | 4.33 |

Note: RL = average remaining life; $t_R$ = optimal preventive replacement; $K_{opt}^{ABR}$ = optimal LCC rate of ABR; $t_I$ = optimal inspection interval; $K_{opt}^{CBR}$ = optimal LCC rate of CBR; $\xi_R$ = optimal preventive replacement threshold.

Firstly, regarding the predicted remaining life (RL) with failure threshold at BCI = 40, the table shows that unlike the conclusion drawn by Giorgio and Pulcini (2024) that the BNGP always underestimate the RL, it may offer overestimation when compared with the BTGP model. Secondly, Bridge #1 exhibits the smallest difference among the three bridges between the two deterioration models for both the ABR and CBR policies. Comparing the 95% predictive bands shown in Figure 6(a) and Figure 12, one can easily find that the BNGP and BTGP models have the closest predictive bands for Bridge #1, particularly in the first 20 to 25 years until BCI reaches 60, which covers the optimized periodic replacement age $t_R$ and the optimized preventive replacement threshold $\xi_R$. In contrast, the predictive bands derived from the two deterioration models differ much greater for the baseline bridge and Bridge #2. Lastly, note that the difference of the mean LCC cost rates between the ABR and CBR policies (i.e., $K_{opt}^{ABR} - K_{opt}^{CBR}$) represents the value of information derived from the inspection. For references of value of information analysis for gamma process, interested readers are referred to (Yuan et al., 2021). In this regard, the BNGP and BTGP provide consistent results for the baseline bridge and Bridge #1, suggesting a positive value of inspection. For Bridge #2, however, the BNGP model suggests preference of ABR policy to the CBR policy, whereas the BTGP does completely the opposite. Note that the derived lifecycle management policies in Table 4 may not be consistent with variations in $\theta_2$, as the comparison is conducted on real bridge historical condition data, rather than a controlled reference group.

*4.3.4 Effect of cost parameters*

The two cost ratios ($C_R/C_I$ and $C_F/C_R$) are important parameters that clearly will affect the optimal ABR and CBR strategies. To further investigate their effects, sensitivity analyses were conducted based on the baseline bridge. Specifically, $C_R/C_I$ varied from 100 to 400 and $C_F/C_R$ varied from 2 to 10. The results are shown in Table 5.

**Table 5 Inspection and replacement results for different cost configurations**

| Case | $C_R/C_I$ | $C_F/C_R$ | BNGP model | | | | | BTGP model | | | | |
|---|---|---|---|---|---|---|---|---|---|---|---|---|
| | | | $t_R$ (yr) | $K_{opt}^{ABR}$ ($/yr) | $t_I$ (yr) | $\xi_R$ | $K_{opt}^{CBR}$ ($/yr) | $t_R$ (yr) | $K_{opt}^{ABR}$ ($/yr) | $t_I$ (yr) | $\xi_R$ | $K_{opt}^{CBR}$ ($/yr) |
| Baseline | 100 | 5 | 44.8 | 2.48 | 6.3 | 47 | 1.91 | 59.2 | 1.76 | 8.5 | 46 | 1.68 |
| Case 1-1 | 200 | 5 | 44.8 | 4.95 | 4.6 | 46 | 3.62 | 59.2 | 3.52 | 5.0 | 44 | 3.21 |
| Case 1-2 | 400 | 5 | 44.8 | 9.90 | 2.8 | 45 | 6.95 | 59.2 | 7.03 | 4.7 | 44 | 6.21 |
| Case 2-1 | 100 | 2 | 54.6 | 2.13 | 6.7 | 45 | 1.79 | 63.3 | 1.66 | 8.6 | 45 | 1.63 |
| Baseline | 100 | 5 | 44.8 | 2.48 | 6.3 | 47 | 1.91 | 59.2 | 1.76 | 8.5 | 46 | 1.68 |
| Case 2-2 | 100 | 10 | 41.1 | 2.67 | 5.0 | 47 | 1.97 | 57.3 | 1.81 | 7.7 | 46 | 1.70 |

With a constant $C_F/C_R$, the optimal age for the ABR is not affected by $C_R/C_I$, whereas the minimized lifecycle cost rate $K_{opt}^{ABR}$ increases proportionally with $C_R/C_I$ because the absolute values of $C_R$ and $C_F$ both have increased proportionally. With a constant $C_R/C_I$, as $C_F/C_R$ increases, the optimal age for the ABR decreases in order to balance the increased failure consequence. As a result, the minimized lifecycle cost rate $K_{opt}^{ABR}$ does not increase proportionally with $C_F/C_R$, suggesting that the ABR strategy is effective in managing the increased failure risk.

For the CBR strategy, a higher $C_R/C_I$ value also indicates cheaper inspections, leading to a decrease in the optimal inspection interval $t_I$. With a constant $C_F/C_R$, the optimal replacement threshold $\xi_R$ decreases as the inspection is done more frequently. With a constant $C_R/C_I$ value in the second group, as $C_F/C_R$ increases the optimal inspection interval $t_I$ decreases as well, although the degree of decrease is much less than the cases of $C_R/C_I$. In addition, an increase in $C_F/C_R$ also causes an increase in the optimal replacement threshold $\xi_R$, suggesting a stricter replacement policy.

Between the BNGP and BTGP models, although these trends apply to both models, the differences of the optimization results are notable across all cost configurations. Therefore, the results above confirm that the two versions of bounded gamma process models may produce significantly different asset management policies.

## 5 Conclusions

Built upon the work primarily presented in Giorgio and Pulcini (2024), the paper proposed a new BTGP model that was deeply rooted on the conventional regression modelling tradition in infrastructure asset management systems. The paper performed both qualitative and quantitative comparisons between the conventional nonstationary gamma process model and the proposed BTGP in terms of predicting future performance and remaining life, as well as designing optimal age- and condition-based replacement policies. An empirical study using a large BCI database was carried out to examine the flexibility of eight candidate models. Our comparisons, together with the comparison presented in Giorgio and Pulcini (2024), made a strong case for the need of a bounded stochastic process model for both deterioration modelling and asset management decision-making. Our empirical studies confirmed that the proposed BTGP model was more flexible than the existing bounded models. Moreover, this study also for the first study identified the MUMV – the mean value at maximum variance – as a critical parameter for determining whether a BTGP model should be used for asset management decision-making.

Most infrastructure performance indicators in infrastructure asset management have an upper bound. This study concludes that a well-designed bounded stochastic process model is not only mathematically possible, but also practically necessary for the sake of conceptual consistency. Moreover, the proposed BTGP is a good candidate for a unified deterioration model for continuous performance measures of various asset classes in a generic infrastructure asset management system.

## Acknowledgements

The financial supports by the Natural Sciences and Engineering Research Council of Canada (NSERC), Infrastructure Canada (RKI-1116) and MTO's 2023-24 Highway Infrastructure Innovation Funding Program (HIIFP) Topic 12 are gratefully acknowledged.

## Declaration of Competing Interest

The authors declare that they have no known competing financial interests or personal relationships that could have appeared to influence the work reported in this paper.